# Quadratic Hierarchy Rule in Flavor Physics and Dual Quark-Neutrino Mixing Patterns

E. M. Lipmanov
40 Wallingford Road # 272, Brighton MA 02135, USA

**Abstract**

Flavor physics is about particle mass-degeneracy-deviations (DMD) and mixing and especially about hierarchies of those deviations. On the one hand there is no established theory of particle flavor at present; on the other hand there are growing data indications on interesting empirical flavor regularities that are described here by two semi-empirical rules – quadratic DMD-hierarchy and Dirac-Majorana DMD-duality rules. First rule unites neutrino solar-atmospheric hierarchy parameter with charged lepton (CL) and quark mass-ratio hierarchies and simultaneously the hierarchies in two mixing matrices of quarks and neutrinos; the second rule predicts quasi-degenerate neutrinos from the fact of CL and quark large mass hierarchy and explains quark-neutrino complementarity mixing relations. The neutrino and quark mixing data seem very different, but it results that small deviations from maximal neutrino mixing are nearly equal to the small deviations from minimal mixing of quarks. Those deviations are quantitatively described by only one new small flavor parameter.

# 1. Introduction

Flavor physics is about elementary particle mass-degeneracy-symmetry deviations and mixing and especially hierarchies of those deviations. The symmetry is rather a reference frame, background, for the definition of the laws of symmetry-violating flavor physics[1]. In [3] and here, on the level of primary phenomenology, I try to solve the flavor problems of small quark mixing versus large neutrino mixing in terms of deviations from extreme, minimal or maximal, mixing values by straight analogy with the lepton deviation-from-mass-degeneracy (DMD) hierarchy approach of refs [1, 2]. A universal quadratic hierarchy equation is formulated in [3] for DMD- and deviation from maximal mixing quantities with lepton mass-degeneracy symmetry essentially violated by suggested DMD-duality[2] relation for the solutions of neutrino and CL hierarchy equation.

In this paper, the small quark mixing parameters are described and interpreted as dual to large neutrino mixing ones with neutrino deviations from maximal mixing replaced by quark deviations from minimal (zero) mixing which obey the same universal quadratic hierarchy rule.

---

[1] A paradigm for the relation between symmetry and its violation in flavor physics can be seen in Newton's classical mechanics: the homogeneous and isotropic absolute space is on the back-ground in classical mechanics; it was needed only as a frame of reference for the definition of the laws of 'symmetry violating' particle motion as the main contents of the theory.

[2] By definition, 'dual' quantities are those which obey the same hierarchy equation and have opposite but connected values.

The absolute numerical values of the mixing parameters for neutrinos and quarks are in plain form expressed through the same universal empirical parameter

$$\alpha_o \cong \exp(-5) \cong 0.0067 \qquad (1)$$

that determines CL and neutrino mass ratios and electroweak interaction constants in [1, 2]. That special value (1) is persistently suggested by very different experimental data.

In Secs.2 the hierarchy equation for extended DMD-quantities is defined and solutions for CL and neutrino mass ratios and deviations from maximal or minimal mixing are obtained. In Sec.3 quantitative results for small quark mixing parameters versus large neutrino mixing ones are discussed. Sec.4 contains conclusions.

## 2. Hierarchy equation for pairs of deviation-from-mass-degeneracy flavor quantities

**I.** The deviation-from-mass-degeneracy quantities of the charged lepton and neutrino mass[3] ratios are described in terms of the new parameter $\alpha_o$ in [1] and [2],

$$(m_\mu^2/m_e^2 - 1) \cong (1/2)(m_\tau^2/m_\mu^2 - 1)^2 \cong 2/\alpha_o^2, \qquad (2)$$

$$(m_2^2/m_1^2 - 1) \cong (1/2)(m_3^2/m_2^2 - 1)^2 \cong 2(5\alpha_o)^2, \qquad (3)$$

and are two solutions of a DMD-hierarchy pattern in lepton flavor physics [1]:

$$(m_\tau^2/m_\mu^2 - 1)^2 / (m_\mu^2/m_e^2 - 1) \cong 2, \qquad (4)$$

$$(m_3^2/m_2^2 - 1)^2 / (m_2^2/m_1^2 - 1) \cong 2. \qquad (5)$$

---

[3] $m_1 < m_2 < m_3$ denote the three neutrino masses.

Eq.(3) for the neutrino DMD-quantities is for the normal neutrino mass ordering; in case of reversed ordering, the ratios $(m_3/m_2)$ and $(m_2/m_1)$ should be interchanged.

In contrast to (2)-(3), the relations (4)-(5) do not depend on any dynamical constant or parameter, are symmetric in structure and may represent two different realizations of a new basic universal flavor hierarchy rule of deviations from symmetry.

That very rule as a universal DMD-hierarchy equation in flavor physics is given by

$$[DMD(2)]^2 \cong 2[DMD(1)], \qquad (6)$$

where DMD(n), n=1,2, denote deviations from unity of the relevant lepton flavor dimensionless quantities: 1) particle mass ratios squared and 2) particle mixing parameters.

By definition, 'symmetric' solution of Eq.(6) is DMD(n) = 0. Its physical meaning in case 1) is exact mass-degeneracy; so, in this case DMD(n) are the regular DMD-quantities $[(m_2/m_1)^2 -1]$, etc, mentioned above. In case 2) DMD(n) are 'extended DMD-quantities' for mixing phenomena. There are two opposite options: deviation from *maximal mixing* DMD(n) = $|\sin^2 2\theta_n -1|$ since DMD(n)=0 would mean $2\theta_n = \pi/2$, and deviation from *minimal mixing* DMD(n)= $|\cos^2 2\theta_n -1|$ since DMD(n)=0 would mean $2\theta_n = 0$.

The *symmetric* solution,

$$DMD(2) = DMD(1) = 0, \qquad (7)$$

means *exact mass-degeneracy* for each of four elementary particle groups – neutrinos, CL and up- and down-quarks. But empirical data definitely disagree with that extreme solution. So, only solutions with deviation from the symmetric one,

$$\text{DMD}(2) = 2a, \quad \text{DMD}(1) = 2a^2, \quad a \neq 0, \tag{8}$$

are possible, they are determined by a real experimental parameter 'a', that measures the symmetry-violation magnitude; its factual value depends on the particular flavor quantity involved. So, in actual flavor physics of known elementary particles the violation of symmetry is always DMD-hierarchical by the quadratic rule (6), or (8). Quadratic hierarchy of lepton flavor physics is considered in [1] for CL and neutrino DMD-quantities and in [3] for neutrino mixing. In quark mixing matrix, quadratic hierarchy is displayed in the Wolfenstein parameterization pattern [10].

The hierarchy rule (6) should answer the specific quantitative neutrino-quark problem of two empirically large solar and atmospheric mixing parameters $\sin^2 2\theta_{12}$ and $\sin^2 2\theta_{23}$ and its *relation to* two *small quark mixing* parameters. I do interpret the hierarchy rule (6) for neutrino and quark mixing parameters in the form:

$$(\sin^2 2\theta_{12} - 1)^2 \cong 2|\sin^2 2\theta_{23} - 1|,$$
$$(\cos^2 2\theta_c - 1)^2 \cong 2|\cos^2 2\theta' - 1|, \tag{9}$$

where $\theta_c$ is the Cabibbo angle and $\theta'$ is the next to the largest quark mixing angle. The DMD(n)-quantities from (6) are interpreted in (9) as *deviations* from maximal or minimal mixing for neutrinos or quarks respectively.

**II)** Consider solutions of Eq.(6) of special interest.

**a.** Large DMD-values $a \gg 1$, CL mass ratios. In this case the DMD-values are large and approximately given by $\text{DMD}(2) \cong (m_\tau/m_\mu)^2 \cong 2a$, $\text{DMD}(1) \cong (m_\mu/m_e)^2 \cong 2a^2$. From comparison [1] with known data, it follows $a \cong 1/\alpha_o$,

$$(m_\tau/m_\mu)^2 \cong 2/\alpha_o, \quad (m_\mu/m_e)^2 \cong 2/\alpha_o^2. \tag{10}$$

The quadratic hierarchy of CL mass ratios is in good agreement with data [1].

   **b.** Small DMD-values $a \ll 1$, QD-neutrinos. With DMD(2) $\cong [(m_3^2/m_2^2) - 1] \cong 2a$, DMD(1) $\cong [(m_2^2/m_1^2) - 1] \cong 2a^2$, this solution means a nearly degenerate mass spectrum and should describe the neutrino mass ratios. Then, the parameter 'a' has a distinct physical meaning being the *observable* in neutrino oscillation experiments solar-atmospheric hierarchy parameter '$r$', and from Eq.(8) we get

$$a \cong (m_2^2 - m_1^2)/(m_3^2 - m_2^2) \equiv r \cong 5\alpha_o \cong 1/30,$$
$$(m_3^2/m_2^2) \cong \exp(2r), \quad (m_2^2/m_1^2) \cong \exp(2r^2). \quad (11)$$

The estimation $r \cong 5e^{-5} = 5\alpha_o \cong 1/30$ in the first line of (11) is from comparison [1] with experimental data.

 With relations (11) the QD-neutrino mass scale is given by

$$m_\nu \cong \sqrt{(\Delta m^2_{atm}/2r)} \cong \sqrt{(\Delta m^2_{sol}/2r^2)}. \quad (12)$$

Using the neutrino oscillation mass-squared differences from data analysis [3-5], the estimation for the QD-neutrino mass scale is around $m_\nu \cong 0.2$ eV, from solar and atmospheric data alike and independent.

   **c.** Large neutrino mixing. With DMD(2) = $|\sin^2 2\theta_{12} - 1| \cong 2a_L \ll 1$, DMD(1) = $|\sin^2 2\theta_{23} - 1| \cong 2a_L^2 \ll 1$, the values in parentheses are *deviations* from maximal mixing [3]. Comparison with experimental *solar* neutrino oscillation data [4, 5, 6] prompts the value $a_L \cong \sqrt{\alpha_o}$,

$$|\sin^2 2\theta_{23} - 1| \cong (1/2)(\sin^2 2\theta_{12} - 1)^2 \cong 2\alpha_o, \quad (13)$$
$$\sin^2 2\theta_{12} \cong \exp(-2\sqrt{\alpha_o}) \cong 0.8486, \quad \theta_{12} \cong 33.6°,$$
$$\sin^2 2\theta_{23} \cong \exp(-2\alpha_o) \cong 0.9866, \quad \theta_{23} \cong 41.7°. \quad (13')$$

The predicted in (13) quadratic hierarchy of deviations from maximal neutrino mixing seems in agreement with data.
-

The choice of the parameter $\alpha_o \ll 1$ (1) in the three pairs of solutions (10) and (11),(13) for the CL and QD-neutrinos is prompted by experimental data. It leads to neutrino-CL *dual DMD-solutions* with very large (CL mass ratios (10)) and very small (QD-neutrino DMD-quantities (11) and small deviations from maximal neutrino mixing (13)) extended DMD-values.

Eq.(13) is a connection between the solar and atmospheric neutrino mixing parameters, that is directly measurable in accurate neutrino oscillation experiments,

$$\cos^2 2\theta_{23} = 0.5 \cos^4 2\theta_{12}. \qquad (14)$$

So, if the large solar neutrino oscillation mixing parameter is not maximal, nonmaximal mixing also follows for the atmospheric oscillation neutrino mixing.

**d).** Small quark mixing: $DMD(2) = |\cos^2 2\theta_c - 1| \cong 2a_Q \ll 1$, $DMD(1) = |\cos^2 2\theta' - 1| \cong 2a_Q^2 \ll 1$, $\theta_c$ is the Cabibbo mixing angle and $\theta'$ is the next to the largest quark mixing angle. Comparison with experimental data of quark mixing [9] prompts a very remarkable inference: there is an approximate, but meaningful equality between the values of the parameters $a_Q$ and $a_L$ for quark and neutrino mixing:

$$a_Q \cong a_L \cong \sqrt{\alpha_o}, \qquad (15)$$

$$\cos^2 2\theta_c \cong \exp(-2\sqrt{\alpha_o}) \cong 0.8486, \quad 2\theta_c \cong 22.9°,$$

$$\cos^2 2\theta' \cong \exp(-2\alpha_o) \cong 0.9866, \quad 2\theta' \cong 6.6°, \qquad (16)$$

in agreement with data, see Sec.3. Quadratic hierarchy of the deviation of the quark mixing parameters from minimal mixing is in agreement with data.

**III)** By the solutions (10) and (13), 'large' neutrino mixing parameters and 'small' quark mixing ones are related to the 'small' charged lepton mass ratios in a symmetric way

$$\sin^2 2\theta_{12} \cong \cos^2 2\theta_c \cong \exp(-2\sqrt{2}\, m_\mu/m_\tau) \cong 0.8452, \qquad (17)$$

$$\sin^2 2\theta_{23} \cong \cos^2 2\theta' \cong \exp(-2\sqrt{2}\, m_e/m_\mu) \cong 0.9864, \qquad (18)$$

in good consistency agreement with (13′) and (16).

**IV)** Relations (17) and (18) indicate a distinct 'reason' of why the neutrino mixing parameters are large, but not maximal – since the charged lepton mass ratios $(m_\mu/m_\tau)$ and $(m_e/m_\mu)$ are small, but not zero (if the electron mass $m_e$ is fixed, zero values for that mass ratios are excluded). And the atmospheric neutrino oscillation angle is closer to maximal than the solar one because of the large empirical CL mass-ratio hierarchy $m_e/m_\mu \ll m_\mu/m_\tau$. By the same reasoning, the quark mixing angles $\theta_c$ and $\theta'$ are different from zero and the Cabibbo angle $\theta_c$ is much larger than $\theta'$. As shown above, the origin of the important here empirical quantitative relation $m_e/m_\mu \ll m_\mu/m_\tau$ is just the quadratic hierarchy rule (6) as for other considered hierarchical generic pairs of flavor quantities.

**V)** The universal DMD-hierarchy rule (6) is independent of $\alpha_o$ and any outer parameter. It should be a primary relation in lepton flavor physics. Lepton flavor physics of the three known flavor generations is probably ruled by *quadratic hierarchy of lepton DMD-quantities* and *neutrino-CL* and neutrino-quark *DMD-duality* which unavoidably violates the mass-degeneracy symmetry. By definition, two dual pairs of quantities are two pairs of quantities which obey the same hierarchy equation but their corresponding members have opposite values and change in opposite directions when the relevant parameter involved is virtually changing. For illustration, in the virtual limit $\alpha_o = 0$ the divergence of CL masses is infinitely large, the neutrinos are exactly mass-degenerate, neutrino mixing is maximal, the divergences of quark mass spectra is infinitely large and quark mixing disappears. In reverse by going from the limit $\alpha_o=0$ to very small actual $\alpha_o$-value (1), one should expect the divergence of CL masses getting finite but large values, the neutrinos getting *quasi*-degenerate, the deviation of neutrino mixing from maximal is small, divergences of quark mass spectra large but finite and the quark mixing parameters getting finite small values.

The condition of DMD-duality predicts 1) QD-neutrinos, 2) small solar-atmospheric hierarchy parameter $r = (m_2^2 - m_1^2)/(m_3^2 - m_2^2) \ll 1$, 3) large solar and atmospheric neutrino mixing parameters and 4) small quark mixing parameters – all four are quantitatively related to the new universal small parameter[4] $\alpha_o$.

As observed in [2], the constant $\alpha_o$ may determine the values of both the fine structure constant $\alpha$ at the photon pole value of momentum transfer and the second electroweak coupling constant $\alpha_W$ at the pole value of the W-boson. So, the new constant $\alpha_o$ should be a quantitative link between flavor physics and the one-generation electroweak physics [12] of leptons and quarks.

---

[4] The common questions of why there are only small and large mixings, but not something in the middle, or why the CL mass spectrum is divergent while the neutrino one is nearly degenerate (if indeed), get convincing answers by the two basic flavor premises 1) extended DMD-duality and 2) one primary universal small flavor-electroweak parameter $\alpha_o$ (1), or its large inverse $1/\alpha_o$, that should determine the magnitudes of all dimensionless flavor quantities.

### 3. Small quark mixing versus large neutrino mixing

I. QD-Majorana neutrinos in flavor physics.

From solutions (13), (16) and (10) of the basic hierarchy equation (6) approximate dual relations follow between neutrino mixing-parameters, quark mixing parameters and CL DMD-quantities,

$$(m_\mu^2/m_e^2 - 1)(1 - \sin^2 2\theta_{23}) \cong 2\sqrt{2}, \qquad (19)$$

$$(m_\tau^2/m_\mu^2 - 1)(1 - \sin^2 2\theta_{12}) \cong 2\sqrt{2}, \qquad (20)$$

$$(m_\tau^2/m_\mu^2 - 1)(1 - \cos^2 2\theta_c) \cong 2\sqrt{2}, \qquad (21)$$

$$(m_\mu^2/m_e^2 - 1)(1 - \cos^2 2\theta') \cong 2\sqrt{2}. \qquad (22)$$

So, the deviations from minimal mixing of quarks are in essence equal to the corresponding deviations from maximal mixing of the neutrinos including the important hierarchies of those deviations.

If the up- and down-quark mass patterns are nearly geometrical and the neutrinos are of Majorana nature, a general form of Dirac-Majorana DMD-duality in flavor physics should be an interesting extension of the considered neutrino-CL DMD-duality [2]. In that case, the neutrino group is a special one in flavor physics, the QD-type of Majorana neutrinos with maximal mixing is contrasted by duality to large mass ratios of Dirac particles, CL and quarks, and small quark mixing.

II. Dual relation between quark and neutrino mixing solutions of the hierarchy equation (6). The relations *between* mixing of quarks and neutrinos can be quantitatively described by the following replacements[5] in the neutrino mixing solution:

$$\{Sin(2\theta_L)\}^{nu} \rightarrow \{Cos(2\theta_q)\}^{quark}, \qquad (23)$$

the superscripts indicate neutrinos and quarks. After such replacement in the neutrino mixing relation (13), the hierarchy equation for the quark mixing parameters is

$$(Cos^2 2\theta' - 1) \cong -(1/2)(Cos^2 2\theta_c - 1)^2 \cong -2\alpha_o, \qquad (24)$$

where $\theta_c$ is the quark largest mixing angle – the *Cabibbo angle* – and $\theta'$ is the next to the largest quark mixing angle. This equation means that the parameters of quark mixing obey the same hierarchy Eq.(6), but in contrast to neutrinos the quark mixing is described by *deviations* from minimal mixing.

III. Wolfenstein hierarchy of the quark mixing parameters.

Rewrite (24) in the form,

$$Sin^2 2\theta_c \cong \sqrt{2} \, Sin \, 2\theta' \cong 2\sqrt{\alpha_o}. \qquad (25)$$

Since both angles $\theta_c$ and $\theta'$ are small, the hierarchy-relation in Eq.(25) is in accord with the Wolfenstein parameterization [10] of the Cabibbo-Kobayashi-Maskawa quark mixing matrix, $Sin \, \theta' \approx (Sin \, \theta_c)^2$ as a reflection of the universal quadratic hierarchy rule (6).

Comparing quark mixing results (25) with neutrino ones (13), one finds interesting duality-like relations between small quark mixing parameters and large neutrino ones

$$Sin \, 2\theta_c \cong Cos \, 2\theta_{12} \cong \sqrt{(2\sqrt{\alpha_o})}, \qquad (26)$$

$$Sin \, 2\theta' \cong Cos \, 2\theta_{23} \cong \sqrt{(2\alpha_o)}. \qquad (27)$$

---

[5] While the quantities $(1 - Sin^2 2\theta_{mix})$ describes deviations from maximal mixing, the other ones $(1 - Cos^2 2\theta_{mix})$ should describe deviations from minimal (zero) mixing.

So, the Cabibbo parameter $\sin 2\theta_c$ is in dual relation to the 'solar' neutrino mixing parameter $\sin 2\theta_{12}$, whereas the next quark mixing parameter $\sin 2\theta'$ is dual to the neutrino 'atmospheric' one $\sin 2\theta_{23}$, and both the quark and neutrino mixing hierarchies come from the universal *quadratic* hierarchy (6).

IV. Smirnov-Raidal quark-neutrino complementarity relations.

From the relation (26) between $2\theta_c$ and $2\theta_{12}$ it follows

$$\sin 2\theta_c \cong \cos 2\theta_{12}, \quad 2\theta_c \cong (\pi/2 - 2\theta_{12}). \quad (28)$$

This relation between the Cabibbo angle and solar neutrino one is already known in quark phenomenology as the quark-lepton complementarity relation [11]. Here it follows from a solution of the universal hierarchy rule (6) and quark-neutrino dual conditions, (26) and (27).

The numerical value of the Cabibbo angle parameter from (26) is given by

$$\theta_c \cong 12°, \quad V_{us} \cong \sin \theta_c \cong 0.21, \quad (29)$$

in agreement with experimental data value [9]

$$(\sin \theta_c)_{exp} \cong 0.22. \quad (30)$$

The next to the largest quark mixing parameter, from (27), is

$$\theta' \cong 3°, \quad V_{cb} \approx \sin \theta' \cong 0.058. \quad (31)$$

In this case, the complementarity relation is given by

$$\sin 2\theta' \cong \cos 2\theta_{23}, \quad 2\theta' \cong (\pi/2 - 2\theta_{23}). \quad (32)$$

So, the empirical quark-neutrino complementarity relations are explained as 'pairs of dual deviation-from minimal or maximal quantities' - solutions of the hierarchy equation (6). Quantitative description of two connected pairs of neutrino and quark mixing parameters on the basis of hierarchy and duality is impressive and do mean unification.

### 4. Conclusions

I. The pair of deviations from maximal mixing of the neutrino oscillation mixing parameters (13) approximately[6] agrees with neutrino oscillation data analysis [4 - 6], and so is an actual test of the discussed lepton flavor extended DMD-hierarchy rule. The other actual independent test is from obtained small mixing of quarks in evident agreement with quark mixing experimental data.

II. The basic equation (6) has a symmetric solution for exact mass-degeneracy of the leptons and quarks see (7). But the subject of flavor physics is rather symmetry *violation,* mainly *hierarchies* of the symmetry violations of generic pairs of flavor quantities[7]. The symmetry (whatever it is[8]) is mainly a frame of reference, a background which is needed for exact definition of the physics laws of symmetry-violation in flavor space. Note that such new approach is straightforward, but *uncommon* in flavor phenomenology since the rules of symmetry violation are here on the fore-ground of the description while the (not detailed) symmetry is only on the back-ground.

---

[6] "…in the description of nature, one has to tolerate approximations, and that even work with approximations can be interesting and can sometimes be beautiful" - P. A. M. Dirac, Scientific autobiography, in *History of 20th Century Physics*, NY (1977).

[7] A generic pair is made of two extended DMD-quantities, which are alike and connected by the quadratic hierarchy equation.

[8] A common feature of any concrete symmetry of that kind is, of course, exact mass-degeneracy of flavor copies.

Dual Dirac-Majorana mass-degeneracy symmetry-*violation* generates QD-neutrino mass spectrum versus divergent CL and quark mass spectra, and also the large neutrino {1-2}- and {2-3}-mixing parameters versus corresponding small quark ones. On the other side, the {1-3}-mixing parameters in neutrino and quark mixing matrices are probably approximately equal [11]. Such suggestion means that in contrast to {1-2}- and {2-3}-mixing parameters the {1-3}-mixing parameters from neutrino and quark mixing matrices are self-dual with magnitude around $\alpha_o/2$ from quark data [9] $s_{13} \cong 0.0036$.

III. QD-neutrino type is still a hypothesis[9]. The established here dual solutions of the hierarchy Eq.(6) are: 1) two pairs of large and small neutrino and quark mixing parameters with hierarchies originated in Eq.(6), 2) two pairs of large CL mass ratios and small deviations from *maximal* neutrino mixing, and 3) two pairs of large CL mass ratios and small deviations from *minimal* quark mixing. If QD-neutrino type will be confirmed by experiment, a fourth pair of dual solutions (DMD-quantities of neutrinos versus CL ones) will be established.

The patterns of large mixing of neutrino mass eigenstates ($\nu_1$, $\nu_2$, $\nu_3$) in the neutrino flavor eigenstates ($\nu_e$, $\nu_\mu$, $\nu_\tau$) and small quark mixing are quantitatively determined by two distinct phenomenological premises – quadratic hierarchy equation and dual relations between generic pairs of solutions - and only one new universal empirical parameter.

---

[9] If further experimental data show that neutrinos are not nearly degenerate in mass or not of Majorana type, the Dirac-Majorana duality interpretation for mass ratios of elementary particles failed, but the quadratic hierarchy and dual relations between large neutrino mixing and small quark mixing should survive.

IV. Plain dependence of all lepton and quark quantitative solutions of Eq.(6) on the same *one small universal parameter* $\alpha_o$ (1) is a significant result persistently suggested by very different lepton and quark experimental data. That one parameter $\alpha_o$ determines and connects all considered dimensionless flavor quantities – two neutrino DMD-quantities (3), solar and atmospheric neutrino mixing parameters (13), two quark mixing parameters (25) and two CL mass ratios (10), see also footnote[10]. In addition, this parameter may also determine the values of the fine structure constant $\alpha$ at zero momentum transfer (photon-propagator pole value) and the second EW interaction constant $\alpha_W$ at pole value of the W-boson propagator [2]. So, the new constant $\alpha_o$ may be a quantitative link between flavor physics and one-generation

---

[10] The eight basic flavor dimensionless quantities (four generic pairs of flavor quantities) related to each other via the universal parameter $\alpha_o$ are given by

$(1/10)\sqrt{(m_2^2/m_1^2 - 1)} \cong (1/10\sqrt{2})(m_3^2/m_2^2 - 1) \cong (1/2\sqrt{2})(\cos 2\theta_{23})^2 \cong$
$(1/4\sqrt{2})(\cos 2\theta_{12})^4 \cong (1/4\sqrt{2})(\sin 2\theta_c)^4 \cong (1/2\sqrt{2})(\sin 2\theta')^2 \cong \sqrt{2}(m_\mu/m_\tau)^2$
$\cong (m_e/m_\mu) \cong \alpha_o/\sqrt{2}$ - in approximate ~(1÷5)% agreement with lepton mass and mixing data and quark mixing data. All that flavor solutions of Eq.(6) are expressed through the constant $\alpha_o$ by comparison with experimental data. Note that the coefficients in these relations get reasonable interpretation in the text.

electroweak physics [12] of leptons and quarks. That result may answer the plain questions of why is the flavor degree of freedom with three particle generations needed at all.

V. Quark and neutrino mixings are empirically very different. Nevertheless, the important physical result is that the *deviations* from maximal neutrino mixing are equal to the *deviations* from minimal quark mixing[11], *including equality of the hierarchies* of those deviations, (19)-(22), and are proportional to the small CL mass *ratios*:

$(1-\cos^2 2\theta_c) \cong (1-\sin^2 2\theta_{12}) \cong 2\sqrt{2}\,(m_\mu/m_\tau)$,

$(1-\cos^2 2\theta') \cong (1-\sin^2 2\theta_{23}) \cong 2\sqrt{2}\,(m_e/m_\mu)$.

VI. Finally, the present phenomenology is supported by different favorable to ideas of hierarchy and duality experimental data especially on neutrino and quark mixing matrices, neutrino oscillation parameters and charged lepton and quark divergent mass spectra and quark-neutrino complementarity.

---

[11] In essence, that result concurs with quark-neutrino complementarity relation by A. Yu. Smirnov and M. Raidal [11].